\documentclass[]{article}

 \usepackage[]{graphicx}
 \usepackage{epsfig}
 \usepackage{amsmath, amstext, amsfonts}
 \usepackage{amssymb}
 \usepackage{multicol}
 \usepackage[latin1]{inputenc}

\usepackage{anysize} 
\marginsize{2cm}{2.5cm}{1.5cm}{2cm}

 \title{Riccati-parameter solutions of nonlinear second-order ODEs}

 \author{M.A. Reyes $^1$ and H.C. Rosu $^2$\footnote{To whom correspondence should be addressed. Email: hcr@ipicyt.edu.mx. Fax: 524448342010. Reyro-08v4.tex}}

 \date{J. Phys. A: Math. Theor. {\bf 41} (2008) 285206 (6pp)\\
 {\small Received 5 March 2008, in final form 13 May 2008\\ Published 19 June 2008}}

\begin{document}

\def\ct{\centerline}
\def\1{\'{\i}}
\def\={\!=\!}
\def\be{\begin{equation}}
\def\ee{\end{equation}}

\maketitle

\begin{center}
$^1$ Instituto de F\'{\i}sica, Universidad de Guanajuato, Le\'on,
Guanajuato, Mexico\\
$^2$ Potos\'{\i} Institute of Science and Technology, Apdo Postal 3-74 Tangamanga, 78231 San Luis Potos\'{\i}, Mexico
\end{center}

\begin{abstract}
\noindent It has been proven by Rosu and Cornejo-P\'erez \cite{rosu1,rosu2} that for
some nonlinear second-order ODEs it is a very simple task to find
one particular solution once the nonlinear equation is factorized
with the use of two first-order differential operators.  Here, it is
shown that an interesting class of parametric solutions are easy to obtain if the proposed factorization has a particular form,
which happily turns out to be the case in many problems of physical
interest. The method that we exemplify with a few explicitly solved cases consists in using the general solution of the Riccati equation,
which contributes with one parameter to this class of parametric solutions.
For these nonlinear cases, the Riccati parameter serves as a `growth' parameter from the trivial null solution up to the particular solution found through the factorization procedure.\\

\end{abstract}

\section{Introduction}

Despite powerful integrability methods, such as the Lie group-theoretical approach, Painlev\'e analysis, existence of Lax representations and the associated inverse scattering transforms, the task of obtaining solutions of nonlinear second order partial and ordinary differential
equations (ODEs) remains one of the most difficult problems in mathematical physics; in some cases, even finding one particular
solution turns out to be a very difficult matter \cite{wang,here}.
However, in a number of cases, it has been proven that finding one
particular solution turns out to be easier than expected. In the case of
polynomial non-linearities, Rosu and Cornejo-P\'erez \cite{rosu1,rosu2}, working with a
factorization procedure stemming from a work of Berkovich
\cite{berko}, have found that if the second order nonlinear
differential equation can be factorized into two first order
differential operators then it is easy to find a first particular
solution for the problem.  They considered a
nonlinear equation of the type
\be\label{nonl} \ddot u + g(u)\,\dot u +F(u)=0~, \ee
where the dots represent derivatives with respect to the independent variable
$\tau$, which is usually the traveling coordinate of a
reaction-diffusion equation \cite{rosu1}. The method they proposed
was to factorize this equation in the following form
\be\label{facRC}
\left[ D-\phi_2(u)\right] \left[ D-\phi_1(u)\right]
u=0~,
\ee
(where $D\equiv \frac{d~}{d\tau}$) which implies the following
conditions on the functions $\phi_i(u)$
\begin{eqnarray}\label{ficond}
-\left( \phi_1+\phi_2+\frac{d\phi_1}{du} \,u \right) &=& g(u)\\
\phi_1\,\phi_2 &=& \frac{F(u)}{u}~.
\end{eqnarray}
If Eq.~(\ref{nonl}) can be factorized as in Eq.~(\ref{facRC}), then a
first particular solution, say $u_1$, can be easily found by solving
\be\label{eq-u1}
\left[ D-\phi_1(u)\right] u=0~.
\ee

\section{Riccati-parameter solutions}

Of course, obtaining one solution of a nonlinear second order ODE
does not guarantee that one may find more general solutions.
However, what Rosu and Cornejo-P\'erez found was that in many cases (some of which will be
described below) the function $\phi_1(u)$ turned out to be a linear
function of the dependent variable $u$.  Hence, Eq.~(\ref{eq-u1})
turns out to be a Riccati equation for this variable, which is very
fortunate, since we already know how to find the general solution
for this equation once a particular solution is known.

The appearance of the Riccati equation in linear second order
differential equations is very common.  In particular, it was very
successfully exploited by Mielnik to find potentials which are
isospectral to the simple harmonic oscillator potential \cite{Bog},
and it is a cornerstone for all SUSY developments \cite{Suk}.
However, it has not been used to solve nonlinear second-order
differential equations at least in the way we present here.

Thus, if $\phi_1$ is of the form $\phi_1(u)=c_1u+c_2$,
Eq.~(\ref{eq-u1}) transforms into the Riccati equation
\be\label{ricc} \dot u -c_1 u^2-c_2u=0 \ee and if a particular
solution $u_1$ of this equation is known, then the general solution
can be found as \cite{Bog}
\be\label{u2}
u_{\lambda, c_1}=u_1+\frac{e^{I_1}}{\lambda-c_1\,I_2}~,
\ee
where
\be\label{i1} I_1(\tau) \equiv \int_{\tau_0}^{\tau}
(2c_1u_1(\tau')+c_2) d\tau'
\ee
and
\be\label{i2}
I_2(\tau)\equiv
\int_{\tau_0}^\tau e^{I_1(\tau')}d\tau' \, .
\ee
Notice that for
\be\label{param}
\lambda _s= c_1 I_2(\tau)~,
\ee
a singularity may develop.

Eq.~(\ref{u2}) provides in
turn what we call as a Riccati-parameter solution of the nonlinear equation. We notice that the first parameter, $c_1$, is essentially the slope of the factorization function $\phi _1$, whereas the $\lambda$
parameter can be chosen in such a way as to prevent this solution from
possessing singularities \cite{Bog}, although for nonlinear differential equations
this is not an absolutely prohibitive issue. It will be seen in the examples given in the following that the latter
parameter acts like a label in this class of solutions placing them between the trivial null solution and the particular solution given by Eq.~(\ref{eq-u1}).

\section{Examples of physical interest}\label{exams}

In this section we find the explicit form of the Riccati-parameter
solution for three nonlinear equations of physical interest that are polynomial type Li\'enard equations, i.e., similar to Eq.~({\ref{nonl}) but with $F(u)$ a polynomial of order two and three in our cases. \\

{\em 3.1 Modified Emden equation}\\

We start with the modified Emden equation
\be\label{emden} \ddot u+\alpha\,u\,\dot u+\beta\, u^3=0~, \ee
for which the first rigorous study has been done by Painlev\'e more than a century ago \cite{P1902} who got solutions for $\beta =\alpha ^2/9$ and
$\beta=-\alpha ^2$.
Recently, Chandrasekar {\it et al} \cite{csl07} provided a detailed discussion of this equation from the point of view of the modified Prelle-Singer procedure that gives the construction of the solution in terms of elementary functions if such a solution exists \cite{chan1}, although
Iacono \cite{iac08} noticed that much simpler connections with the Abel equation could be used to get the solutions. For the remarkable physical applications, see \cite{csl07}.

Employing $\phi_1=a_1\sqrt{\beta}\,u$ and
$\phi_2=a_1^{-1}\sqrt{\beta}\,u$, where $a_1=-\frac{\alpha\pm
\sqrt{\alpha^2-8\beta}}{4\sqrt\beta}$, one particular solution is
\cite{rosu2}
\be\label{emdenu1}
u_1= -\frac{1}{a_1\sqrt\beta (\tau-\tau_0)}~.
\ee
Hence, by using Eq.~(\ref{u2}) one can find that the two-parameter
solution of Eq.~(\ref{emden}) is
\be\label{emdenu2}
u_{\lambda}= -\frac{1}{a_1\sqrt\beta
(\tau-\tau_0)}+ \frac{1}{\lambda(\tau-\tau_0)^2 +
a_1\sqrt\beta(\tau-\tau_0)}~.
\ee
In this case it is instructive to notice that when $|\lambda|$ runs
from zero to infinity, the $u_{\lambda}$ solution goes from the trivial
solution $u=0$ to the particular solution $u=u_1$, as can be deduced from Eq.~(\ref{emdenu2}) and graphically seen in
figure \ref{fig1}.\\

\begin{figure}[htb]
\centerline{
\includegraphics[height=6.5cm, width=8cm]{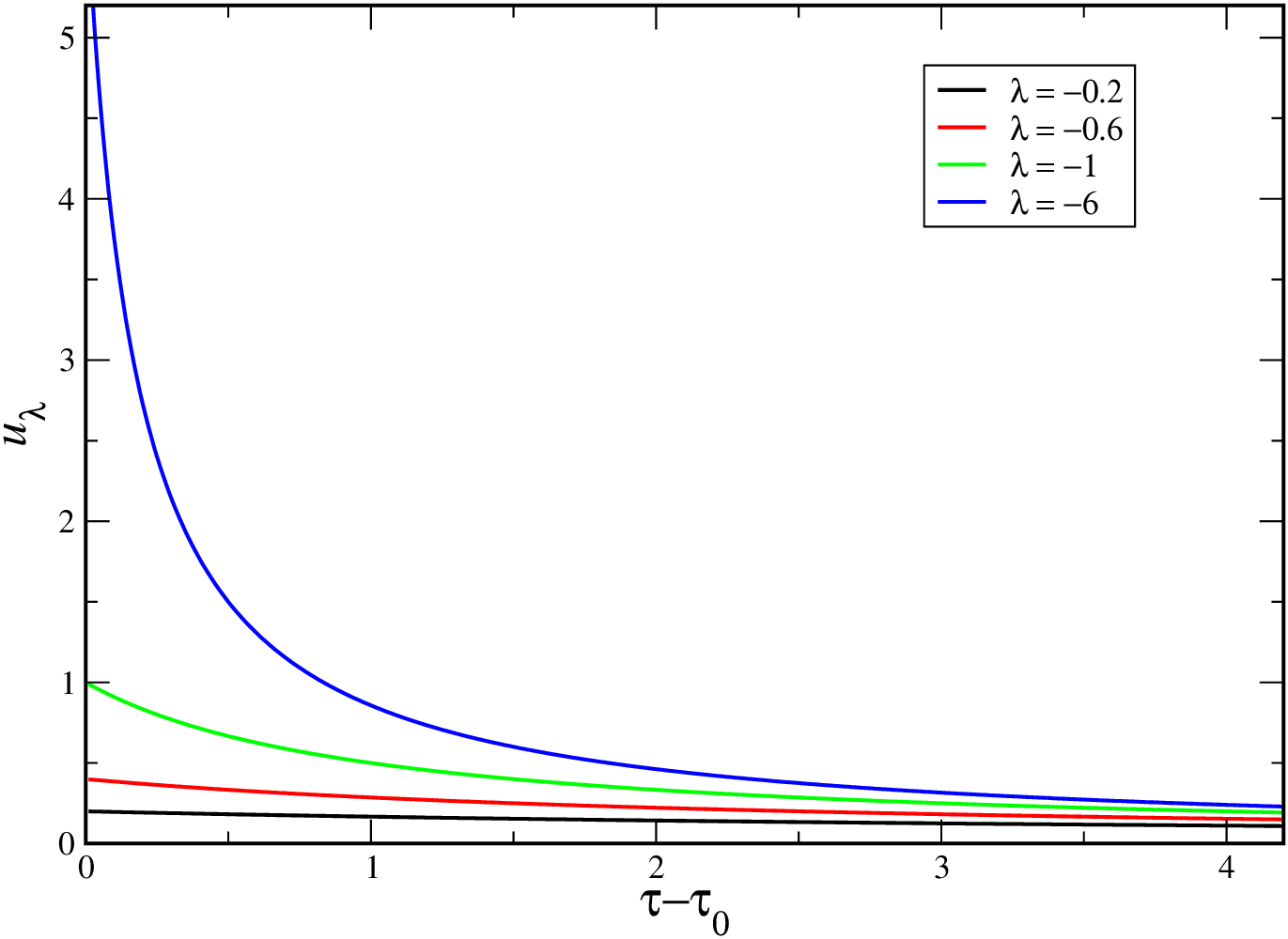}}
\caption{$u_{\lambda}(\tau)$ in the case of the modified Emden equation for
$\lambda=-0.2,-0.4,-1,-6$, from top to bottom respectively, and $a_1\sqrt\beta=-1$.} \label{fig1}
\end{figure}

{\em 3.2 Convective Fisher equation}\\

We pass now to the convective Fisher equation \cite{schon}
\be \ddot
u+2(\nu-\mu\,u)\dot u +2u(1-u)=0~.
\ee
The second term corresponding to convection is introduced to describe mechanical transport in competition with diffusive transport or cases when
external bias fields are present. In the context of population dynamics which is typical for the Fisher equation, it has been used by Walsh {\em et al} \cite{walsh} to simulate the population mobility according to spatial gradients in the food supply.\\

Rosu and Cornejo-P\'erez found that for
$\nu=\mu/2+\mu^{-1}$, the factorization functions are $\phi_1=-\mu(1-u)$ and $\phi_2=-2/\mu$. Hence, a
particular solution for this equation will be \cite{rosu2}
\be\label{fishu} u_1=\left[ 1\pm \exp\left( \mu (\tau-\tau_0)
\right) \right]^{-1}~. \ee
The $\lambda$-parameter solution can be readily obtained in the form
\be\label{fishu2}
u_{\lambda} = u_1+ 
\frac{e^{-\mu(\tau-\tau_0)}}
{\left[ e^{-\mu(\tau-\tau_0)} \pm 1 \right]
\left[ \lambda\left(e^{-\mu(\tau-\tau_0)}\pm 1\right)-1 \right]}~.
\ee
Once again, $u_{0}=0$ and $u_{\infty}=u_1$. For the graphical representation see figure \ref{fig2}.\\

\begin{figure}[htb]
\centerline{
\includegraphics[height=6.5cm, width=8cm]{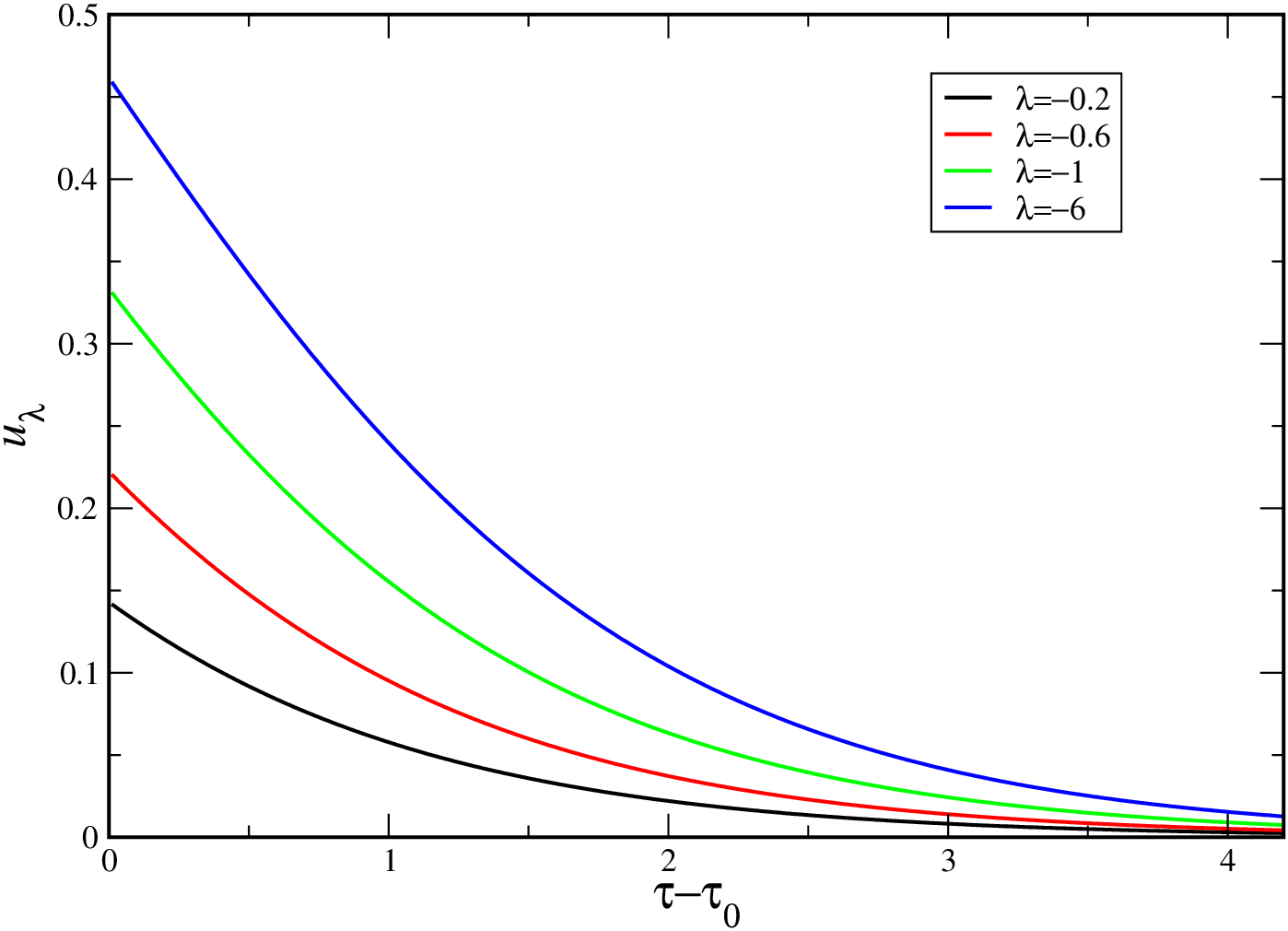}}
\caption{Convective Fisher $u_{\lambda}(\tau)$ for
$\lambda=-0.2,-0.4,-1,-6$, from top to bottom respectively, and $\mu=1$.} \label{fig2}
\end{figure}

\bigskip

{\em 3.3 Generalized Li\'enard equation}\\

Consider now the generalized Li\'enard equation for cubic nonlinear oscillators
\be\label{glien}
\ddot u+g(u)\,\dot u +F_3=0~, \ee where $F_3(u)=Au+Bu^2+Cu^3$.
The previous equations can be seen as particular cases of this one.
With $\Delta=\sqrt{B^2-4AC}$, Rosu and Cornejo-P\'erez found that using
\[
\phi_1=a_1\left( \frac{B+\Delta}{2}+Cu \right) \ \ , \ \ \ \
\phi_2=a_1^{-1}\left( \frac{B-\Delta}{2C}+u \right)
\]
for
\be
g(u)= -\left[
\frac{B+\Delta}{2}a_1+\frac{B-\Delta}{2C}a_1^{-1}+
\left( 2Ca_1+a_1^{-1} \right)u
\right]
\ee
the following particular solution could be obtained \cite{rosu2}
\be\label{glienu}
u_1(\tau)=\frac{\frac{B+\Delta}{2}}{\exp\left(
-a_1\frac{B+\Delta}{2} (\tau-\tau_0) \right)
-C}=\frac{\frac{B+\Delta}{2}\exp\left( a_1\frac{B+\Delta}{2}
(\tau-\tau_0)\right)}{1 -C\exp\left( a_1\frac{B+\Delta}{2}
(\tau-\tau_0) \right)}~.
\ee
Now, using Eq.~(\ref{u2}) and denoting $\xi=\tau -\tau _0$, we can see
that the two-parameter solution in this case is
\begin{equation}\label{glienu2}
u_{\lambda}(\xi) = u_1+
\frac{\frac{B+\Delta}{2}\exp\left(-a_1\frac{B+\Delta}{2}\xi\right)}
{\left[\exp\left( -a_1\frac{B+\Delta}{2}\xi\right)-C\right]
\left[\left(\lambda\frac{B+\Delta}{2}-1\right)\exp\left(-a_1\frac{B+\Delta}{2} \xi\right)
-\lambda C\frac{B+\Delta}{2}\right]}~.
\end{equation}
Varying $\lambda$ between 0 and $\infty$, the Riccati-parameter solutions go from the null solution to $u_1$.
Plots of $u_{\lambda}$ for several values of $\lambda$ are displayed in figure \ref{fig3}.\\


%
\begin{figure}[htb]
\centerline{
\includegraphics[height=5.5cm, width=8cm]{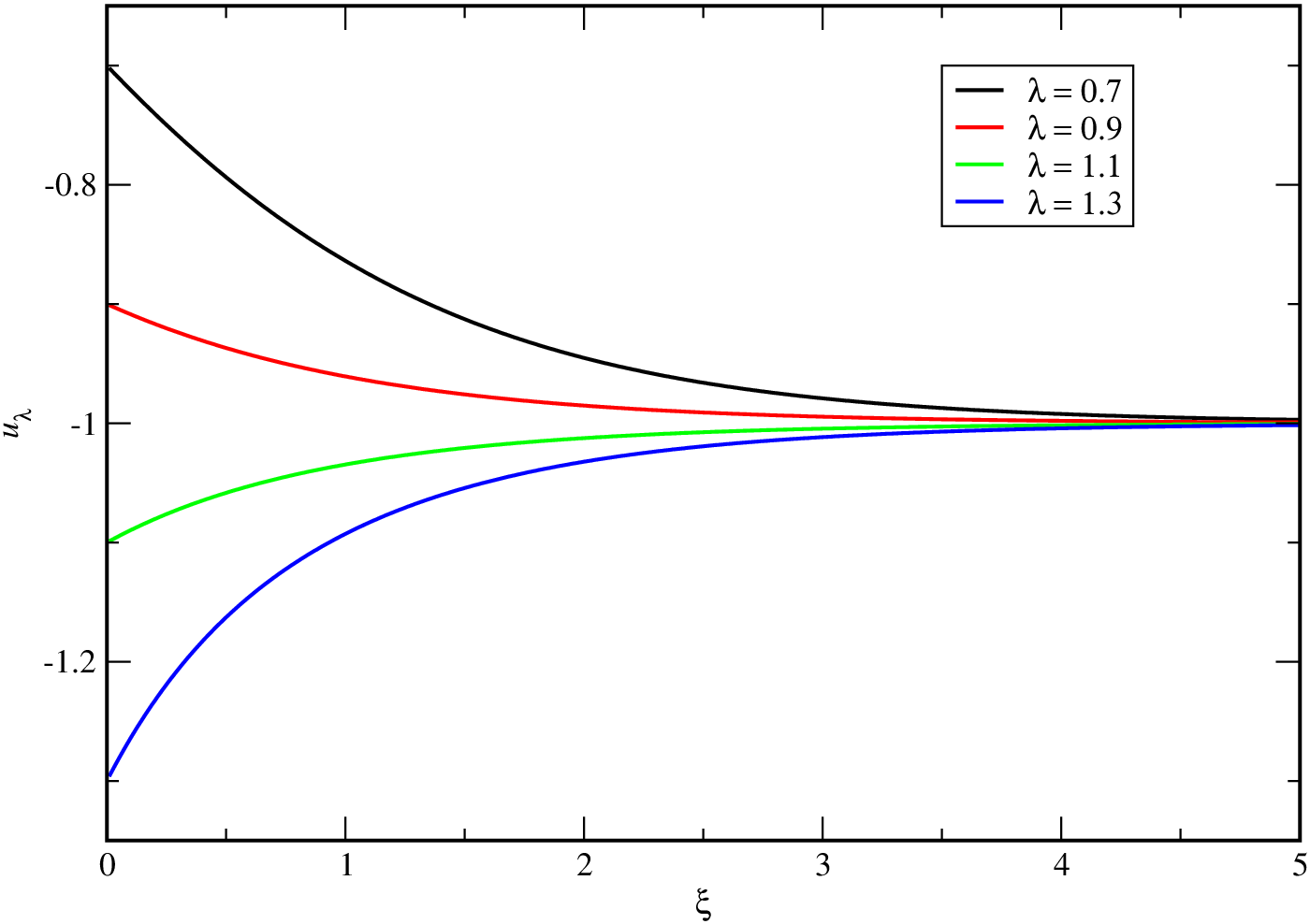}} 
\caption{Cubic Li\'enard $u_{\lambda}(\xi)$ solutions for
$\lambda=0.7,0.9,1.1,1.3$, from top to bottom respectively, in the case $A=C=1$, $B=2$ ($\Delta =0$) and $a_1=1$.} \label{fig3}
\end{figure}

\bigskip

{\em 3.4 Other cases of physical interest}\\

The examples we have provided here are not the only possible cases
that can be solved with this method, but they show the typical
solutions to be found. Other examples where Riccati-parameter solutions
can be obtained in this way are: the generalized Burgers-Huxley equation with
$\delta=1,\ \gamma=1$, the generalized Fisher equation,
with $n=2$, the Dixon-Tuszynski-Otwinowski type
equation, with $n=4$, and the Fitzhugh-Nagumo equation
\cite{rosu1}. The linear factorization functions $\phi _1$ of all these cases
are given in \cite{rosu1}. Last but not least, we would like to comment on the possible physical interpretation of the Riccati parameter $\lambda$.
We follow the works of Barton {\em et al} \cite{barton90} and Monthus {\em et al} \cite{monthus96} to assert that $\lambda$ is related to the introduction of finite interval boundaries on the abscissa of the problem. This is very well described in Section II A of \cite{monthus96} to which the interested reader is directed.
Essentially, the introduction of boundary conditions at certain points on the axis generates a modulation of the particular solution as presented here
and the $\lambda$ parameter can be fixed through the boundary conditions. When the boundary is sent to infinity the original particular solution is
recovered.

\bigskip

In conclusion, we introduced here an interesting class of parametric solutions of a number of physically relevant nonlinear differential
equations. They cover the space between the null or constant solution and the particular solution obtained by a simple factorization method proposed previously.

\bigskip

We would like to thank Dr. O. Cornejo-P\'erez for a careful reading of the first draft of this work.
The second author wishes to thank CONACyT for partial support through
project 46980.

\end{document}